# Oxygen Point Defect Chemistry in Ruddlesden-Popper Oxides $(La_{1-x}Sr_x)_2MO_{4\pm\delta}$ (M = Co, Ni, Cu)

*Wei Xie[1§], Yueh-Lin Lee[1,2†], Yang Shao-Horn[2], Dane Morgan[1\*]*

[1]Department of Materials Science and Engineering, University of Wisconsin–Madison, Madison, Wisconsin 53706, United States

[2]Electrochemical Energy Laboratory, Massachusetts Institute of Technology, 77 Massachusetts Avenue, Cambridge, Massachusetts 02139, United States

**Corresponding Author**

[*]D.M.: Email: ddmorgan@wisc.edu.

**Present address**

[§]W.X.: University of California, Berkeley, Berkeley, CA 94720, United States

[†]Y.L.L.: National Energy Technology Laboratory, Pittsburgh, PA 15236, United States

## ABSTRACT

Stability of oxygen point defects in Ruddlesden-Popper oxides $(La_{1-x}Sr_x)_2MO_{4\pm\delta}$ (M = Co, Ni, Cu) is studied with density functional theory calculations to determine their stable sites, charge states, and energetics as functions of Sr content ($x$), transition metal (M) and defect concentration ($\delta$). We demonstrate that the dominant O point defects can change between oxide interstitials, peroxide interstitials, and vacancies. Generally, increasing $x$ and atomic number of M stabilizes peroxide over oxide interstitials, as well as vacancies over both peroxide and oxide interstitials; increasing $\delta$ destabilizes both oxide interstitials and vacancies, but affects little peroxide interstitials. We also demonstrate that the O 2$p$-band center is a powerful descriptor for these materials and correlates linearly with the formation energy of all the defects. The trends of formation energy versus $x$, M and $\delta$ and the correlation with O 2$p$-band center are explained in terms of oxidation chemistry and electronic structure.

**TOC GRAPHICS**

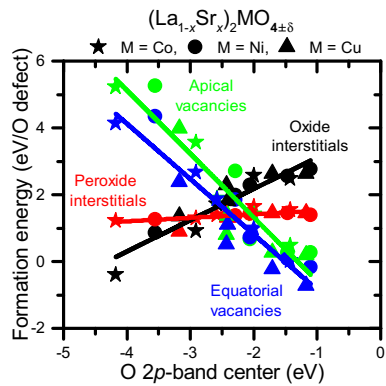

Ruddlesden-Popper (RP) oxides with the $K_2NiF_4$ structure ($RP_{214}$), for example $(A_{1-x}A'_x)_2MO_{4\pm\delta}$ (A = La, Pr, Nd; A' = Ca, Sr, Ba; M = Co, Ni, Cu) are promising potential materials for solid oxide fuel cell (SOFC) electrodes and ion transport membranes due to their being excellent mixed electronic and ionic conductors (MEICs) and having high surface reactivity with $O_2$[1-2]. $RP_{214}$ oxides can have both O vacancy and interstitial as active O point defects, although under temperature $T$ and oxygen partial pressure $P(O_2)$ relevant to typical conditions of SOFC and similar applications, many $RP_{214}$ oxides are believed to have O interstitials as the majority O defect[3-4]. Despite the widespread interest in $RP_{214}$ materials, there is still significant uncertainly about their defect chemistry. In particular, the nature of O interstitials—whether in oxide $O^{2-}$ or peroxide $O^{1-}$ state—is controversial[5-12]. While it appears that O interstitial may exist in both states, their relative stability and changes with the content of aliovalent dopant $x$, transition metal element M and O defect concentration $\delta$ are not well understood. Furthermore, the widely accepted mechanism for oxide interstitial oxygen transport effectively passes through the peroxide during the interstitialcy mechanism hop, suggesting that the relative energetics of these interstitial states may play a major role in interstitial kinetics.[13-14] Overall, the balance of the oxygen interstitial and vacancy energetics controls defect chemistry and in turn determines the associated material properties like electrical conductivity[15-16], magnetism[16-17], and oxygen surface exchange[15, 18] and bulk transport[19-20].

Besides defect stability itself, determining how electronic structure controls O defect stability and correlates with the O surface exchange and bulk transport kinetics is important to enhance understanding and enable rational design and optimization of these materials through both computation and experiment. Very recently, Nakamura et al.[21] have studied $La_2NiO_4$ based $RP_{214}$ oxides with soft X-ray absorption spectroscopy and proposed that unoccupied projected density of states (pDOS) of the 2p electrons of lattice O is the determining factor for interstitial O formation in $La_2NiO_4$ based $RP_{214}$ oxides. Similarly, Lee et al.[22] have extended previous studies on perovskites[23] and reported that the computed mass centroid of both the occupied and unoccupied pDOS of 2p electrons of lattice O atoms relative to the Fermi level (often simply referred to as O 2p-band center) is an electronic structure descriptor that correlates with calculated bulk and surface O defect formation and absorption energies as well as measured activation energies for O surface exchange and bulk transport for selected $RP_{214}$ oxides.

However, this initial work[22] leaves a number of open questions critical for the design of oxygen transport devices. In particular, the nature of the defect chemistry, including under what compositions different materials might be dominated by interstitial vs. vacancy O point defects and to what extent this defect chemistry can be robustly correlated with O 2p-band center, is still not established. Furthermore, a unified understanding in terms of changes in oxidation state and electronic structure to guide thinking across the $RP_{214}$ family of materials needs to be developed.

Here we report a density functional theory study of the stability of oxygen point defects in a series of $RP_{214}$ oxides $(La_{1-x}Sr_x)_2MO_{4\pm\delta}$ (M = Co, Ni, Cu) with Sr mole fraction covering the whole range ($x$ = 0, 0.25, 0.5, 0.75, 1) and three O defect concentrations ($\delta$ = 0.0625, 0.125, 0.25). Four types of O point defects are studied including O vacancies at the apical and equatorial positions of the $MO_2$ octahedra, as well as O interstitials in the oxide $O^{2-}$ and peroxide $O^{1-}$ states, which are all illustrated in Figure S1 (Supporting Information). Each defect is modeled by either adding or removing one O atom from the relaxed pristine supercells of bulk $(La_{1-x}Sr_x)_2MO_4$, leading to either $(La_{1-x}Sr_x)_2MO_{4+\delta}$ or $(La_{1-x}Sr_x)_2MO_{4-\delta}$ (i.e., $\delta$ is always positive). Details of the computational approach and its validation against experiments[24-30] are provided in the Supporting Information. This work will address the following key questions. First, what are the relative stabilities (as characterized by the formation energy, $E_{def}$) of the various O point defects in bulk $(La_{1-x}Sr_x)_2MO_{4\pm\delta}$ under typical conditions of SOFC and similar applications (e.g., $T = 1000$ K, $P(O_2) = 0.21$ atm)? Second, what are trends of O defect formation energy $E_{def}$ vs. Sr mole fraction $x$, transition metal element M, and O defect concentration $\delta$ and how can they be interpreted? Third, to what extent does $E_{def}$ correlate with O 2$p$-band center for all the four types of O point defects over the full range of relevant $x$, $\delta$ and M, and what is the nature of these correlations? Answering these questions provides quantitative energies, qualitative trends, physical understanding and electronic structure descriptor for the defect chemistry of many of the RP$_{214}$ oxides most active to oxygen surface exchange and bulk transport, thereby laying a foundation for their uses in SOFC electrodes and other active oxygen applications.

Figure 1 presents O point defect formation energy $E_{def}$ versus Sr mole fraction $x$. The same formation energy is plotted versus transition metal M and O defect concentration $\delta$ in Figure S3 and Figure S4 (Supporting Materials), respectively. A few important properties of of defect stability are immediately apparent from these figures. First, we find that equatorial O vacancies are always more or at least equally stable than apical vacancies for all the cases of different M, $x$ or $\delta$ considered here, consistent with previous experimental and computational results for LaSrCo$_{0.5}$Fe$_{0.5}$O$_4$[32] and La$_{1.85}$Sr$_{0.15}$CuO$_4$[33]. Second, unlike vacancies, the relative stability of the two O interstitial states depends significantly on M and $x$ and less on $\delta$. O interstitials are in general more or at least equally stable in the oxide state compared to the peroxide state on the La end (i.e., $x = 0$), except for the one case of M= Cu with $\delta = 0.25$ (the bottom right panel), but are always more stable in the peroxide state on the Sr end (i.e., $x = 0$). In the middle, the formation energy curves for the two interstitial states cross, which occurs at smaller $x$ as the atomic number of M or the defect concentration $\delta$ increases. For example, when $\delta = 0.125$ the crossings are near $x$ = 0.50, 0.20 and 0 for M = Co, Ni, Cu, respectively; for M = Ni, the crossings are near $x$ = 0.25, 0.20 and 0.12 when $\delta$ = 0.0625, 0.125, and 0.25 respectively. Finally, comparing the most stable O vacancies with the most stable O interstitials, Figure 1 shows that under the typical SOFC condition of $T$ = 1000 K and $P(O_2)$ = 0.21 atm the former are further more stable than the latter towards the Sr rich end while the opposite is true towards the La rich end. The formation energies of the two again cross in the middle, near $x$ = 0.50 for M = Co, 0.25 for M = Ni, and 0.10 for M = Cu when $\delta$ = 0.0625. As $\delta$ further increases these crossovers occur at larger $x$ (slightly larger $x$ when $\delta$ = 0.125 and substantially larger $x$ when $\delta$ = 0.25). Table S3 (Supporting Information) further gives some more precise comparisons to experimentally measured crossovers[27-30], showing discrepancies between the model and experiment less than 0.1 in $x$ for all the three systems. At higher $T$ or lower $P(O_2)$ (i.e., more reducing conditions), the O chemical potential becomes smaller, so O interstitials will become less stable while vacancies become more so, and hence the crossover will occur at smaller $x$. Notably, towards the La rich end peroxide interstitials can indeed be the most stable O point defects in $(La_{1-x}Sr_x)_2MO_{4\pm\delta}$ especially for M = Ni and Cu and when $\delta$ is sufficiently large (e.g., $\delta$ = 0.125, which may be reached in O rich environment), consistent with previous experimental results[5, 11].

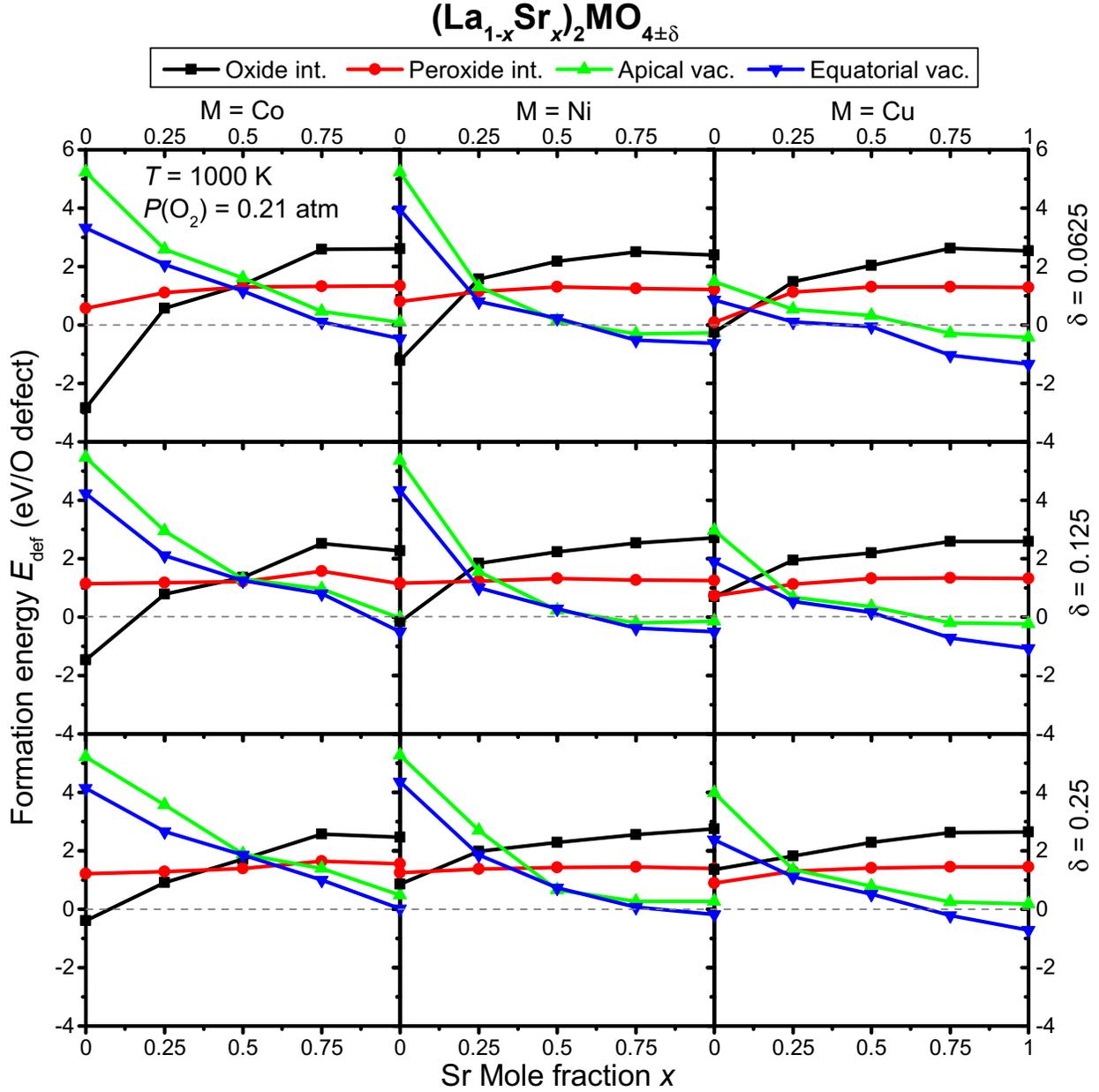

**Figure 1**. Formation energy for O point defects versus Sr mole fraction $x$ in $(La_{1-x}Sr_x)_2MO_{4\pm\delta}$. The referenced O chemical potential corresponds to $T = 1000$ K and $P(O_2) = 0.21$ atm (Supporting Information)[31]. The three columns are for M = Co, Ni and Cu while the three rows are for $\delta = 0.0625$, 0.125 and 0.25, respectively.

Overall, the formation energies generally increase or are flat for oxide interstitials while decrease or are flat for both equatorial and apical vacancies with higher Sr content $x$ (Figure 1) and larger atomic number of transition metal M (Figure S3). The effects of both $x$ and M on the formation energies become smaller at larger $x$, in particular for oxide interstitials. For example, Figure 1 (Figure S3) shows that formation energy curves of oxide interstitials seem to level out with $x$ (with M) approximately after $x \geq 0.75$. In contrast to their opposite trends with $x$ and M, the formation energies for oxide interstitials and vacancies (both equatorial and apical) are generally both increased with larger O defect concentration $\delta$ (Figure S4). For peroxide interstitials, however, the defect stability generally does not seem to be affected by $x$, M, or $\delta$, except to some small extents when $x \approx 0$.

All of these trends can be readily understood in terms of the oxidation chemistry of the defect formation and the electronic structure of $RP_{214}$ oxides. Increasing $x$ oxidizes the system, which effectively lowers the Fermi level relative to vacuum, as does increasing the atomic number M from Co to Ni and further to Cu. With a lower Fermi level, oxidative defects like the oxide interstitial become less stable while reductive defects like the vacancies become more so. The peroxide interstitial is largely unaffected by changes in the Fermi level (and therefore $x$ and M) as they form by the reaction $O^{2-}(solid) + 1/2O_2(gas) \rightarrow O_2^{2-}(solid)$, which does not involve any redox in the oxides.

The reduced sensitivity of oxide interstitials to $x$ and M when $x$ is larger than some threshold occurs because oxide interstitials are trying to oxidize a system whose transition metal cannot be easily oxidized further. That is, transition metal ions are in oxidation states with prohibitively high oxidation energy (i.e., the energy required going to a higher oxidation state). As a result, oxide interstitials will oxidize lattice O rather than transition metal ions and therefore the formation energy of oxide interstitials becomes largely independent of Sr content, as the lattice O ions change relatively little with increasing Sr content due to the larger number of available O atoms and associated electrons compared to transition metals. The proceeding argument is supported by the experimentally measured average oxidation number for Co ions in $(La_{1-x}Sr_x)_2CoO_{4\pm\delta}$[27], which increases with Sr content initially but levels out for larger Sr content. Similarly, once oxygen redox rather than transition metal redox is playing a dominant role we can also expect the formation energy of oxide interstitials to be relatively insensitive to the transition metal M, which explains the level out of $E_{def}$ vs. M after $x \geq 0.75$ observed in Figure S3 (Supporting Information). Therefore, it is the oxidation of the system to a level where redox is effectively pinned by oxygen redox (vs. transition metal redox) that leads to a reduced sensitivity to $x$ and M.

The trends of $E_{def}$ versus $\delta$ can also be understood in terms of oxidation and electronic structure changes, although here we must also consider mechanisms of direct defect interaction such as electrostatics. In general, we would expect defects to become less stable with increasing $\delta$ due to strain, electrostatics, and Fermi level effects, which is what is observed in our calculations and consistent with experiments (e.g., the increase in formation energy with defect concentration $\delta$ has also been observed experimentally[34] for $(La_{1-x}Sr_x)_2NiO_{4+\delta}$). Furthermore, because the redox active defects (oxide interstitials and vacancies) have more significant electrostatic interactions and change the Fermi level in a manner that makes them harder to form, they are expected to have stronger $\delta$ dependence, which is also generally what is observed. However, precise trends with $\delta$ can depend significantly on many factors, including electrostatic and strain interactions between the oxygen defects as well as the specific defect and Sr locations, and thus a more quantitative analysis of factors governing formation energy must treat specific cases rather than all those cases studied here.

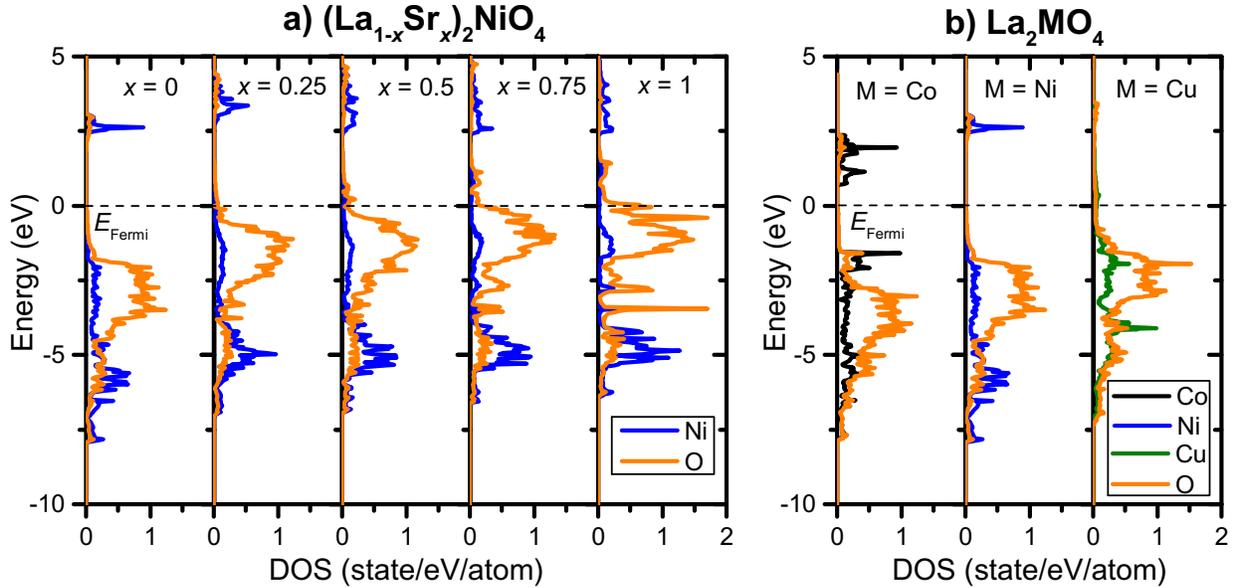

**Figure 2**. Electronic density of states for a) $(La_{1-x}Sr_x)_2NiO_4$ with $x$ = 0, 0.25, 0.5, 0.75 and 1, and b) $La_2MO_4$ with M = Co, Ni and Cu.

The redox arguments given above to explain the trends of $E_{def}$ versus $x$ and M can be verified by examining the electronic density of states (DOS) for our materials, which are provided in Figure 2. Figure 2 a) illustrates the trends with $x$ using the DOS of $(La_{1-x}Sr_x)_2NiO_4$. When $x$ increases from 0 to 0.75, the pDOS of transition metal M moves down relative to oxygen which effectively brings down the overall Fermi level to be closer to, and eventually dominated by, the O pDOS, with the transition appearing to be complete near $x$ = 0.75. The shift in the metal M bands makes the system harder to oxidize by oxide O interstitials and easier to reduce by O vacancies at higher $x$. The relative change (e.g., between $x$ = 0 - 0.25 and $x$ = 0.25 – 0.5) in the energies of the M bands becomes smaller when $x$ is larger, explaining why the effect of further increasing $x$ is smaller when $x$ is larger. Going further from $x$ = 0.75 to 1 then appears to only create holes in the oxygen states (i.e., oxidizing the lattice oxygen), and shifts the Fermi level by only a very small amount. Note that here we have only shown the case with M = Ni. For those with other transition metals, the trends are similar, but the threshold $x$ may be different.

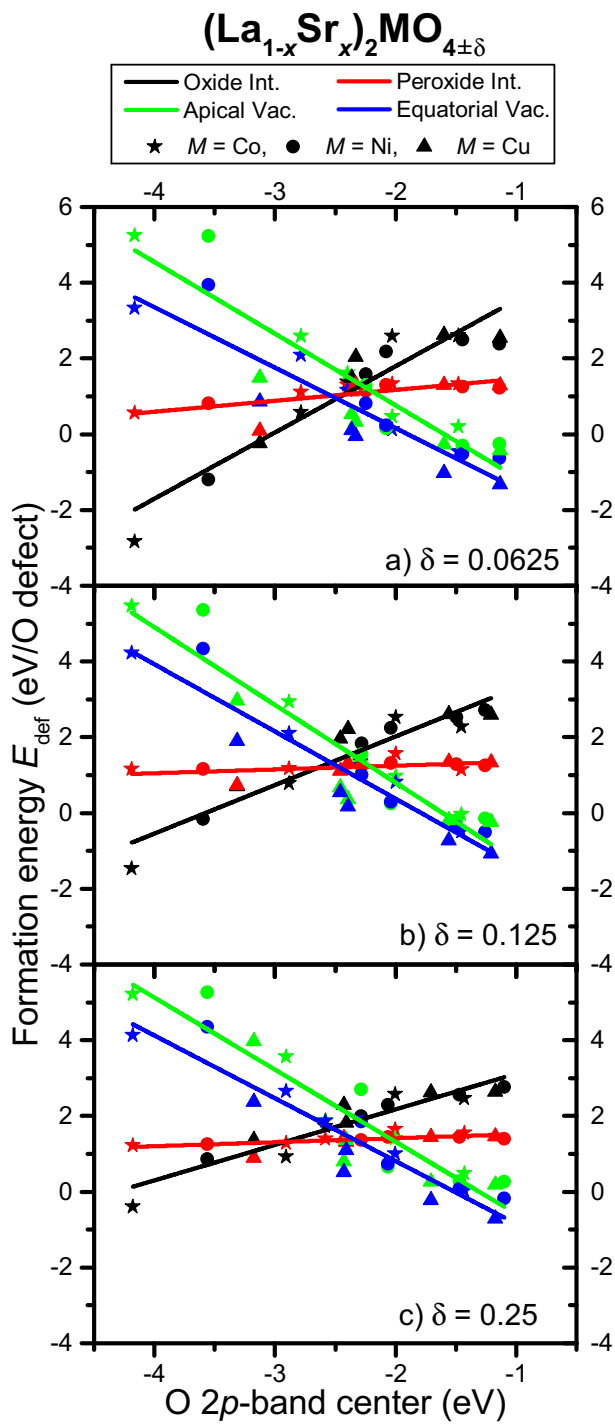

Figure 3. Formation energy of O point defects versus O 2$p$-band center (relative to the Fermi level) in bulk $(La_{1-x}Sr_x)_2MO_{4\pm\delta}$ (M=Co, Ni, Cu) with $\delta$ = a) 0.0625, b) 0.125, and c)

**0.25**. The referenced O chemical potential μ(O) corresponds to $T = 1000$ K and $P(O_2) = 0.21$ atm (Supporting Information)[31]. For each type (color) of defect, star, circle and triangle symbols indicate actually calculated values for M = Co, Ni and Cu, respectively, while the straight line is a linear fitting of them. Details of the fittings are provided in Table S4 (Supporting Information).

The trends with M can be explained using the DOS of $La_2MO_4$ in Figure 2 b). As the atomic number M increases, the metal M pDOS is lower in energy with respect to the O pDOS, corresponding to the higher metal electronegativity and associated redox energy. Thus substituting M with M' of higher atomic number has the same qualitative effect as increasing $x$. This result explains why M ions become harder to oxidize by oxide O interstitials but easier to reduce by O vacancies when the transition metal M changes from Co to Cu. Note that here we have only shown the case with Sr content $x = 0$. Although not shown in Figure 2, we note that with larger $x$, the transition metal M bands will have smaller difference in the energy and eventually overlap when $x$ is over some threshold, explaining the decreasing effect of M when $x$ becomes larger.

Figure 3 shows that formation energy of all the four O point defects in bulk $(La_{1-x}Sr_x)_2MO_{4\pm\delta}$ correlates linearly with their O 2p-band centers. As given in Table S4 (Supporting Information), the coefficients of determination ($R^2$) are between 0.8 and 0.9 for all defects except the peroxide interstitials, for which the near-zero slopes lead to low $R^2$ despite the good linear correlation. Outliers for the fittings are those for O vacancies at $x = 0$ in the M = Ni case with O 2p-band center near -3.5 eV (not labeled in Figure 3), which are also those breaking the linear $E_{def}$ vs. M trend mentioned above. We believe these outliers are due to the fact that vacancy formation in $La_2MO_4$ reduces M from 2+ to 1+, which can lead to physics distinct from all the other defect formation reactions studied here as they include only 2+/3+/4+ defect states. The 2+/1+ redox is unphysical for Co and Ni, and perhaps even Cu, and therefore not of significant interest so we make no effort to address these outliers. The slopes for the fitting curves are almost 0, positive and negative for peroxide interstitials, oxide interstitials, and vacancies, respectively. These trends are as expected, as higher 2$p$-band center corresponds to a system that is harder to oxidize and easier to reduce. These linear relationships can be used for rapid computational screening as the O 2$p$-band descriptor can be determined much faster than the defect energies. Furthermore, the result that defect formation energies in the RP$_{214}$ phases depend linearly on O 2$p$-band suggests that this descriptor will be successful for describing many of the oxygen transport and catalytic properties of RP$_{214}$ phases, as already demonstrated for activation energies by Lee, *et al.*[22]

The linear relationship with the O 2$p$-band can again be interpreted based on electronic structure. The O 2$p$-band center relative to vacuum is determined by stability of electrons in lattice O atoms, which in turn are determined largely by the O electron affinity and electrostatic Madelung potential. We propose that the latter two are little changed by altering $x$ and M within the RP$_{214}$ structure, and that the O 2$p$-band center is therefore an approximately fixed reference vs. vacuum. Within this approximation the O 2$p$-band center relative to the Fermi energy is also a measure of the changes in the Fermi level relative to a fixed vacuum. To the extent that most of the changes in defect energies with $x$, M, and $\delta$ are dominated by changes in redox energy they will be approximately linearly related to changes in the Fermi level relative to vacuum, and therefore linearly related to the O 2$p$-band center (relative to the Fermi level). These correlations then explain the linear relationship between $E_{\text{def}}$ and O 2$p$-band center. This explanation further suggests that the slopes should correlate to the number of electrons involved in the redox of the defect, which is in fact what we observed, as discussed in the Supporting Information.

In summary, we identified the stable point defects for the RP 214 oxides $(La_{1-x}Sr_x)_2MO_{4\pm\delta}$ (M = Co, Ni, Cu) including the composition leading to changes in the dominant defects from interstitials to vacancies and from oxide ($O^{2-}$) to peroxide ($O^{1-}$) interstitials. Under the typical SOFC condition of $T$ = 1000 K and $P(O_2)$ = 0.21 atm, O interstitials are preferred over O vacancies when no Sr is doped, become similarly stable as O vacancies in the middle (e.g., near $x$ = 0.50 for M = Co, 0.25 for M = Ni, and 0.10 for M = Cu when $\delta$ = 0.0625), and are dominated by O vacancies afterwards. Most importantly, O interstitials in the peroxide state can indeed be the most stable O point defects in $(La_{1-x}Sr_x)_2MO_{4\pm\delta}$ with small Sr doping content $x$ especially for M = Ni and Cu and when $\delta$ is sufficiently large (e.g., $\delta$ = 0.125). Furthermore, we showed that formation energy of all the four studied O point defects in bulk $(La_{1-x}Sr_x)_2MO_{4\pm\delta}$ correlate linearly with their O 2$p$-band center. We also explained the trends with $x$, M, $\delta$ and O 2$p$-band center in terms of changing redox state of the system and relate them to the changing electronic structure.

The understanding developed in this work may help guide the design and use of $RP_{214}$ phases for applications involving oxygen related catalytic processes and oxygen transport, including fuel cell components and ion exchange membranes. Furthermore, our results give insights into the rather novel area of oxygen redox, a topic of increasing interest in developing redox active systems[35]. Finally, the identification of the O 2$p$-band center as a robust descriptor across different M and $x$ may simplify the interpretation of O defect related materials properties and computational screening of $RP_{214}$ oxides. More broadly, given that O vacancy formation energy in $ABO_3$ perovskite ($PV_{113}$) oxides also correlates linearly with their O 2$p$-band centers[23], we believe linear correlation between formation energy of O point defects and O 2$p$-band center may be a general phenomenon across many classes of oxides, a hypothesis currently being studied further.

## ACKNOWLEDGMENTS

Initial calculations for this work by Y.-L. Lee were supported by the U.S. Department of Energy (DOE), National Energy Technology Laboratory (NETL), Solid State Energy Conversion Alliance (SECA) Core Technology Program with award number FE0009435. The bulk of the calculations, all manuscript development, and all activities by W. Xie and D. Morgan were supported by the NSF Software Infrastructure for Sustained Innovation (SI$^2$) award No. 1148011. This work used the Extreme Science and Engineering Discovery Environment (XSEDE), which is supported by U.S. National Science Foundation with grant number ACI-1053575.

## ASSOCIATED CONTENT

**Supporting Information**

Details of the computational approach and validation against experiment, all numerical results of the formation energy, effects of Sr/La arrangement and defect locations, formation energy versus transition metal M and O defect concentration δ, and more details of the linear fitting of formation energy versus O 2$p$-band center.

**Notes**

The authors declare no competing financial interest.

## REFERENCES

(1).	Jacobson, A. J. Materials for Solid Oxide Fuel Cells. *Chem. Mater.* **2010,** *22*, 660-674.
(2).	Bouwmeester, H. J. M. Dense Ceramic Membranes for Methane Conversion. *Catal. Today* **2003,** *82*, 141-150.
(3).	Skinner, S. J.; Kilner, J. A. Oxygen Diffusion and Surface Exchange in La2-xSrxNiO4+delta. *Solid State Ionics* **2000,** *135*, 709-712.

(4).    Munnings, C. N.; Skinner, S. J.; Amow, G.; Whitfield, P. S.; Davidson, I. J. Oxygen Transport in the La2Ni1-xCoxO4+delta System. *Solid State Ionics* **2005,** *176*, 1895-1901.

(5).    Buttrey, D. J.; Ganguly, P.; Honig, J. M.; Rao, C. N. R.; Schartman, R. R.; Subbanna, G. N. Oxygen Excess in Layered Lanthanide Nickelates. *J. Solid State Chem.* **1988,** *74*, 233-238.

(6).    Jorgensen, J. D.; Dabrowski, B.; Pei, S.; Richards, D. R.; Hinks, D. G. Structure Of the Interstitial Oxygen Defect in La2NiO4+Delta. *Phys. Rev. B* **1989,** *40*, 2187-2199.

(7).    Chaillout, C.; Chenavas, J.; Cheong, S. W.; Fisk, Z.; Marezio, M.; Morosin, B.; Schirber, J. E. 2-Phase Structural Refinement of La2CuO4.032 at 15 K. *Phys. C* **1990,** *170*, 87-94.

(8).    Radaelli, P. G.; Jorgensen, J. D.; Schultz, A. J.; Hunter, B. A.; Wagner, J. L.; Chou, F. C.; Johnston, D. C. Structure of the Superconducting La2CuO4+delta Phases (delta-Approximate-to-0.08,0.12) Prepared by Electrochemical Oxidation. *Phys. Rev. B* **1993,** *48*, 499-510.

(9).    Demourgues, A.; Weill, F.; Darriet, B.; Wattiaux, A.; Grenier, J. C.; Gravereau, P.; Pouchard, M. Additional Oxygen Ordering in La2NiO4.25 (La8Ni4O17) .1. Electron and Neutron-Diffraction Study. *J. Solid State Chem.* **1993,** *106*, 317-329.

(10).    Wu, Y.; Ellis, D. E.; Shen, L.; Mason, T. O. Point Defects of La2CuO4-Based Ceramics .1. Oxygen Interstitials. *J. Am. Ceram. Soc.* **1996,** *79*, 1599-1604.

(11).    Li, Z. G.; Feng, H. H.; Yang, Z. Y.; Hamed, A.; Ting, S. T.; Hor, P. H.; Bhavaraju, S.; DiCarlo, J. F.; Jacobson, A. J. Carrier-Controlled Doping Efficiency in La2CuO4+delta. *Phys. Rev. Lett.* **1996,** *77*, 5413-5416.

(12).    Cordero, F.; Grandini, C. R.; Cantelli, R. Structure, Mobility and Clustering of Interstitial O in La2CuO4+delta in the Limit of Small delta. *Phys. C* **1998,** *305*, 251-261.

(13).    Chroneos, A.; Parfitt, D.; Kilner, J. A.; Grimes, R. W. Anisotropic Oxygen Diffusion in Tetragonal La2NiO4+delta: Molecular Dynamics Calculations. *J. Mater. Chem.* **2010,** *20*, 266.

(14).    Li, X.; Benedek, N. A. Enhancement of Ionic Transport in Complex Oxides through Soft Lattice Modes and Epitaxial Strain. *Chem. Mater.* **2015,** *27*, 2647-2652.

(15).    Garcia, G.; Burriel, M.; Bonanos, N.; Santiso, J. Electrical Conductivity And Oxygen Exchange Kinetics of La2NiO4+delta Thin Films Grown By Chemical Vapor Deposition. *J. Electrochem. Soc.* **2008,** *155*, P28-P32.

(16).    Greenblatt, M. Ruddlesden-Popper Ln(n+1)Ni(n)O(3n+1) Nickelates: Structure and Properties. *Curr. Opin. Solid State Mater. Sci.* **1997,** *2*, 174-183.

(17).    Rodriguezcarvajal, J.; Fernandezdiaz, M. T.; Martinez, J. L. Neutron-Diffraction Study on Structural and Magnetic-Properties of La2NiO4. *J Phys.: Condens. Matter* **1991,** *3*, 3215-3234.

(18).    Tsipis, E. V.; Naumovich, E. N.; Shaula, A. L.; Patrakeev, M. V.; Waerenborgh, J. C.; Kharton, V. V. Oxygen Nonstoichiometry and Ionic Transport in La2Ni(Fe)O4+delta. *Solid State Ionics* **2008,** *179*, 57-60.

(19).    Kharton, V. V.; Tsipis, E. V.; Yaremchenko, A. A.; Frade, J. R. Surface-Limited Oxygen Transport and Electrode Properties of La(2)Ni(0.8)Cu(0.2)O(4+delta). *Solid State Ionics* **2004,** *166*, 327-337.

(20).    Kovalevsky, A. V.; Kharton, V. V.; Yaremchenko, A. A.; Pivak, Y. V.; Naumovich, E. N.; Frade, J. R. Stability and Oxygen Transport Properties of Pr2NiO4+delta Ceramics. *J. Eur. Ceram. Soc.* **2007,** *27*, 4269-4272.

(21).    Nakamura, T.; Oike, R.; Ling, Y.; Tamenori, Y.; Amezawa, K. The Determining Factor for Interstitial Oxygen Formation in Ruddlesden-Popper Type La2NiO4-Based Ixides. *Phys. Chem. Chem. Phys.* **2016,** *18*, 1564-9.

(22). Lee, Y.-L.; Lee, D.; Wang, X. R.; Lee, H. N.; Morgan, D.; Shao-Horn, Y. Kinetics of Oxygen Surface Exchange on Epitaxial Ruddlesden–Popper Phases and Correlations to First-Principles Descriptors. *J. Phys. Chem. Lett*. **2016,** *7*, 244-249.

(23). Lee, Y. L.; Kleis, J.; Rossmeisl, J.; Shao-Horn, Y.; Morgan, D. Prediction of Solid Oxide Fuel Cell Cathode Activity with First-Principles Descriptors. *Energ Environ. Sci*. **2011,** *4*, 3966-3970.

(24). Prasanna, T. R. S.; Navrotsky, A. Energetics of La2-xSrxCoO4-y (0.5-Less-Than-x-Less-Than-1.5). *J. Solid State Chem*. **1994,** *112*, 192-195.

(25). Dicarlo, J.; Mehta, A.; Banschick, D.; Navrotsky, A. The Energetics of La2-xAxNiO4-Y (A = Ba, Sr). *J. Solid State Chem*. **1993,** *103*, 186-192.

(26). Bularzik, J.; Navrotsky, A.; Dicarlo, J.; Bringley, J.; Scott, B.; Trail, S. Energetics of La2-XSrxCuo4-Y Solid-Solutions (0.0-Less-Than-or-Equal-to-X-Less-Than-or-Equal-to-1.0). *J. Solid State Chem*. **1991,** *93*, 418-429.

(27). Nitadori, T.; Muramatsu, M.; Misono, M. Valence Control, Reactivity of Oxygen, and Catalytic Activity of lanthanum Strontium Cobalt Oxide (La2-xSrxCoO4). *Chem. Mater*. **1989,** *1*, 215-220.

(28). Takeda, Y.; Kanno, R.; Sakano, M.; Yamamoto, O.; Takano, M.; Bando, Y.; Akinaga, H.; Takita, K.; Goodenough, J. B. Crystal-Chemistry and Physical-Properties of La2-xSrxNiO4 (0 <= x <= 1.6). *Mater. Res. Bull*. **1990,** *25*, 293-306.

(29). Kanai, H.; Mizusaki, J.; Tagawa, H.; Hoshiyama, S.; Hirano, K.; Fujita, K.; Tezuka, M.; Hashimoto, T. Defect Chemistry of La2-xSrxCuO4-delta: Oxygen Nonstoichiometry and Thermodynamic Stability. *J. Solid State Chem*. **1997,** *131*, 150-159.

(30). Opila, E. J.; Tuller, H. L. Thermogravimetric Analysis and Defect Models of the Oxygen Nonstoichiometry in La2-XSrxCuO4-Y. *J. Am. Ceram. Soc*. **1994,** *77*, 2727-2737.

(31). Lee, Y. L.; Kleis, J.; Rossmeisl, J.; Morgan, D. Ab initio Energetics of LaBO3(001) (B=Mn, Fe, Co, and Ni) for Solid Oxide Fuel Cell Cathodes. *Phys. Rev. B* **2009,** *80*, 224101.

(32). Tomkiewicz, A. C.; Tamimi, M.; Huq, A.; McIntosh, S. Oxygen Transport Pathways in Ruddlesden-Popper Structured Oxides Revealed via in situ Neutron Diffraction. *J Mater. Chem. A* **2015,** *3*, 21864-21874.

(33). Meyer, T. L.; Jiang, L.; Lee, J.; Yoon, M.; Freeland, J. W.; Jang, J. H.; Aidhy, D. S.; Borisevich, A.; Chisholm, M.; Egami, T.; et al.. Selective Control of Oxygen Sublattice Stability by Epitaxial Strain in Ruddlesden-Popper films. *arXiv* **2015**, *1508*, 06971*,* arXiv.org e-Print archive. http://arxiv.org/abs/1508.06971 (accessed February 2016).

(34). Nakamura, T.; Yashiro, K.; Sato, K.; Mizusaki, J. Oxygen Nonstoichiometry and Defect Equilibrium in La2-xSrxNiO4+delta. *Solid State Ionics* **2009,** *180*, 368-376.

(35). Grimaud, A.; Hong, W. T.; Shao-Horn, Y.; Tarascon, J. M. Anionic Redox Processes for Electrochemical Eevices. *Nat. Mater*. **2016,** *15*, 121-126.

# Oxygen Point Defect Chemistry in Ruddlesden-Popper Oxides $(La_{1-x}Sr_x)_2MO_{4\pm\delta}$ (M = Co, Ni, Cu)

*Wei Xie$^{1\S}$, Yueh-Lin Lee$^{1,2\dagger}$, Yang Shao-Horn$^2$, Dane Morgan$^{1*}$*

$^1$Department of Materials Science and Engineering, University of Wisconsin–Madison, Madison, Wisconsin 53706, United States

$^2$Electrochemical Energy Laboratory, Massachusetts Institute of Technology, 77 Massachusetts Avenue, Cambridge, Massachusetts 02139, United States

**Corresponding Author**

$^*$D.M.: Email: ddmorgan@wisc.edu.

**Present address**

$^\S$W.X.: University of California, Berkeley, Berkeley, CA 94720, United States

$^\dagger$Y.L.L.: National Energy Technology Laboratory, Pittsburgh, PA 15236, United States

## S1: COMPUTATIONAL METHODS

Each system of $(La_{1-x}Sr_x)_2MO_{4\pm\delta}$ (M=Co, Ni, Cu) was modeled with three different supercell sizes of the same tetragonal Bravais lattice: $\sqrt{2}a \times \sqrt{2}a \times c$, $2a \times 2a \times c$, $2\sqrt{2}a \times 2\sqrt{2}a \times c$, where $a$ and $c$ are the lattice vectors parallel and perpendicular to the rock salt/perovskite layers, respectively of the 14-atom conventional unit cell of the $K_2NiF_4$ structure, often denoted high temperature tetragonal (HTT) (space group 139, *I*4/*mmm*). Therefore, the pristine supercells have 28, 56 and 112 atoms, respectively. Although the Bravais lattice of the supercells was kept tetragonal to better simulate the HTT structure that is usually the most stable structure under conditions relevant to SOFC and related applications, coordinates of atoms were taken from the ground state structure, often denoted low temperature orthorhombic (LTO) (space group 64, *Bmab*) as initial atomic positions for relaxation. Namely, the $MO_2$ octahedra were already manually tilted as in LTO. Otherwise tilting may happen when relaxing the atom positions of defected supercells, which will introduce unphysical change to defect formation energy. That is to say, Bravais lattice of the lattice (i.e., the parallelepiped box) of the simulation supercells is tetragonal, but the Bravais lattice of the crystal (i.e., lattice + asymmetric unit) of the simulation supercells is orthorhombic. For pristine supercells, atom positions, volume and $c/a$ ratio were relaxed while the lattice was kept tetragonal. Thus we explored the energetics of the fully relaxed structure with octahedral tilting subject to the constraint that Bravais lattice of the lattice of the simulations supercells is kept tetragonal.

Based on relaxed pristine supercells, defected supercells were created by either adding or removing one O atom, and the resulted defected $\sqrt{2}a \times \sqrt{2}a \times c$, $2a \times 2a \times c$, and $2\sqrt{2}a \times 2\sqrt{2}a \times c$ supercells have O defect concentration $\delta$ = 0.25, 0.125, and 0.0625, respectively. Note we write the general chemical formula for defected supercells as $(La_{1-x}Sr_x)_2MO_{4\pm\delta}$. For a supercell containing one vacancy, it should be $(La_{1-x}Sr_x)_2MO_{4-\delta}$, while for a supercell containing one interstitial, it should be $(La_{1-x}Sr_x)_2MO_{4+\delta}$. This way, $\delta$ is always positive. Apical and equatorial O vacancies were created by removing one lattice O atom from the apical and equatorial position of one $MO_2$ octahedron in the $(La_{1-x}Sr_x)O$ rock salt and $(La_{1-x}Sr_x)MO_3$ perovskite layer, respectively. O interstitial in oxide ($O^{2-}$) and peroxide ($O^{1-}$) state were created by adding one O atom, both in the rock salt layer, but about 2.6 and 1.5 Å apart from an apical lattice O atom, respectively. Wyckoff symbols for oxide and peroxide interstitials should be 4$d$ and 16$n$, respectively in terms of HTT's space group, while typical atomic coordinates for oxide and peroxide interstitials are roughly (0.75, 0.25, 0.25) and (0.63, 0.14, 0.29) in terms of the $\sqrt{2}a \times \sqrt{2}a \times c$ supercell lattice. The resulted defected supercells are visualized in Figure S1. Only atom positions were relaxed for the defected supercells and the cell parameters were held constant. Note that sometimes during relaxation an interstitial O atom may move from an initial oxide interstitial position to that of peroxide interstitial (as could be judged by the distance to the nearest apical lattice O), or the other way around. Checking, and if necessary re-relaxation were therefore performed to ensure the results reported here indeed correspond to the expected interstitial states.

Each system of $(La_{1-x}Sr_x)_2MO_{4\pm\delta}$ (M=Co, Ni, Cu) was modeled at five different mole fractions of Sr: $x$ = 0, 0.25, 0.5, 0.75, 1. When La or Sr fully occupy the A-site, as occurs at $x$ = 0 or 1, all possible positions for each defect also belong to the same Wyckoff site and are symmetrically equivalent. However, at $x$ = 0.25, 0.5 and 0.75, the symmetry is lowered as La and Sr are mixed, leading to two complications. First, the A site can be occupied by La and Sr in multiple distinct ways. Second, for each of these A-site orderings, the previously equivalent defect positions become symmetrically distinct, creating many possible defect positions for each type of O defect. We treated this problem by assuming randomly disordered La and Sr and arranging them using the special quasi-random structure (SQS) method[1] as implemented in the Alloy Theoretic Automated Toolkit (ATAT)[2-3]. We used two and three different $\sqrt{2}a \times \sqrt{2}a \times c$ SQS structures at $x = 0.5$ and $x = 0.25/0.75$, respectively. For each SQS we enumerated all distinct defect positions. The resulted defect formation energies are provided in Figure S2. It is encouraging that the spread in defect formation energies due to different La/Sr arrangement and defect positions, despite being sizable in some cases (particularly for O vacancies), does not seem to blur the effects of Sr doping $x$ and transition metal M, which produce quite consistent trends in Figure S1 that are obvious even with the spread. These trends are the main phenomena we hope to explore in this study, not that due to La/Sr arrangement. We found the lowest energy configurations were not always the same for different transition metal M. As a result, we picked one typical low energy set of SQS structure and defect positions and used them throughout all the three systems of M = Co, Ni and Cu. Finally, we generated the $2a \times 2a \times c$ and $2\sqrt{2}a \times 2\sqrt{2}a \times c$ defected structures as supercells of the selected $\sqrt{2}a \times \sqrt{2}a \times c$ defected structures, keeping still only one O defect per unit cell. All the INCAR, POSCAR, KPOINTS and CONTAR files, which include the final relaxed coordinates for all the defect calculations used in this study, are also given as digital supporting data.

All calculations were performed in the general framework of spin-polarized Density Functional Theory (DFT)[4-5] using the Vienna *Ab initio* Simulation Package (VASP)[6-7] version 5.3.3 based on the projector-augmented-wave (PAW) method[8] as implemented by Kresse and Joubert[9]. The exchange-correlation functional was treated in the generalized gradient approximation (GGA) following Perdew and Wang[10] (PW-91). On-site Hubbard $U$ potential was added for $d$-electrons of the transition metal elements in the simplified rotationally invariant DFT + $U$ approach[11] with effective Hubbard $U$ parameter $U_{eff}$ = 3.3, 6.4 and 4.0 eV for Co, Ni and Cu, respectively[12]. The PAW potentials used for La, Sr, Co, Ni, Cu, and O were labeled by the VASP developers as La 14Apr2000, Sr_sv 10Feb1998, Co 03Mar1998, Ni_pv 19Apr2000, Cu 05Jan2001, O_s 04May1998, which treated $5s^26p^65d^16s^2$, $4s^24p^65s^2$, $3d^84s^1$, $3p^63d^94s^1$, $3d^{10}4p^1$ and $2s^22p^4$ as valence electrons, respectively. The stopping criteria for self-consistent loops were 0.1 meV/cell and 1 meV/cell tolerance of total energy for the electronic and ionic relaxation, respectively. A plane wave cutoff energy of 425 eV was used. The partial occupancies were set using Gaussian smearing with a smearing width of 0.2 eV. The electronic and ionic optimizations were performed using the Davidson-block algorithm[13] and the Conjugate-gradient algorithm[14], respectively. The projection operators were evaluated in reciprocal space and non-spherical contributions from the gradient corrections inside the PAW spheres were included. Monkhorst–Pack[15] $k$-point meshes of $6\times6\times3$, $4\times4\times2$, and $3\times3\times2$ were used to sample the Brillouin zone for the $\sqrt{2}a\times\sqrt{2}a\times c$, $2a\times2a\times c$, $2\sqrt{2}a\times2\sqrt{2}a\times c$ supercells, respectively.

In order to use a consistent and tractable set of magnetic structures, all calculations were performed in the ferromagnetic state to simulate a magnetic environment more similar to the application relevant high-temperature paramagnetic state than the typically antiferromagnetic ground state[16]. Disordered moments in paramagnetic state are significantly more difficult to model and we believe that the trends and conclusions identified here would not be altered by using random spin arrangements. For M = Co, the relaxed spin states seemed to evolve from high spin to intermediate spin with increasing Sr mole fraction $x$, but were in general very difficult to reach convergence as we found that VASP's optimization algorithm often became stuck in metastable spin states. Therefore, we also manually searched for the ground spin states using series of calculations with constrained magnetic moments. The current results are the lowest energy solutions from both VASP's optimization and our manual searches. We believe for most cases the spin states should have been converged. For a few selected cases (e.g., at $x$ = 0.75, M = Co, $\delta$ = 0.125 in the first panel of second row in Figure 1), convergence may not have been reached despite our extensive efforts, which leads to small kinks in the formation energy, $E_{def}$, but does not affect the main trends of $E_{def}$ vs. $x$, M, and $\delta$ that are focus of this study. On the other hand, VASP's optimization algorithm usually had no difficulty reaching convergence for M = Ni and Cu. The relaxed spin states are low spin for M = Ni. For M = Cu, the $d$ shell is mostly full without distinction of spin states, and the relaxed magnetic moments are typically very small or zero.

The O 2$p$-band center was determined using the mass centroid of the projected O 2$p$-states (including both the occupied and unoccupied states) relative to Fermi level of relaxed pristine supercells. The formation energy $E_{def}$ for O point defects was calculated as:

$$E_{def}(T, P(O_2)) = E((La_{1-x}Sr_x)_2MO_{4\pm\delta}) - E((La_{1-x}Sr_x)_2MO_4) \mp \delta\mu_{O_2}(T, P(O_2)),$$

where $E((La_{1-x}Sr_x)_2MO_4)$ and $E((La_{1-x}Sr_x)_2MO_{4\pm\delta})$ are the total energy of the defected supercell $(La_{1-x}Sr_x)_2MO_{4\pm\delta}$ and pristine supercell $(La_{1-x}Sr_x)_2MO_4$, respectively, both from *ab initio* calculations at 0 K in this study, while $\mu_{O_2}(T, P(O_2))$ is gas phase O chemical potential. We used $\mu_{O_2}(T, P(O_2))$ values that were estimated according to Equation (3) of Ref.[16]. More specifically, for PW-91 GGA based calculations, $\mu_{O_2}^{DFT} = \frac{1}{2}[E_{O_2}^{VASP} + \Delta h_{O_2}^0]$ = -8.76 eV/O$_2$ = -4.38 eV/(O atom) is just the corrected O$_2$ molecule total energy[16]; adding to it the change in the free energy of O$_2$ from the reference state $(T^0, P^0(O_2))$ = (298.15 K, 1 atm), which we obtained from Ref.[17], $\mu_{O_2}$ at finite $T$ and $P(O_2)$ was estimated, for example $\mu_{O_2}(1000\,K, 0.21\,atm)$ = -5.59 eV/(O atom). No charge compensation was performed (i.e., no additional charge was added, so the defected supercells were neutral overall although defects could have localized charges) as all the systems are metallic except perhaps those with $x = 0$.

## S2: Validation of computation approach against experiment

All results of O defect formation energy referenced to just the corrected $O_2$ molecule total energy $\mu_{O_2}^{DFT}$ = -4.38 eV/(O atom)[16] are tabulated in Table S1. Based on them, results at finite $T$ and $P(O_2)$ can be obtained using corresponding O chemical potentials (see Section S2 above). Table S2 compares our computed formation energy with experimental results we find in the literature[18-20]. The discrepancy is < 0.4 eV/(O defect), which is typical for defect energetics due to experimental uncertainty, DFT errors, possible differences in the computational setup from experiments, and the simplified identification of our formation enthalpy with that measured in full defect chemistry of the experiment. Table S3 further compares the critical $x$ where $(La_{1-x}Sr_x)_2MO_{4\pm\delta}$ switches between O hypo- and hyper-stoichiometric, which is determined experimentally by equilibrium O contents[21-24]. We estimate this crossover from our data as where the calculated most stable O interstitial energy curve crosses with that of the most stable O vacancy. We find the discrepancy between the calculation and the experiment to be less than 0.1 for all three systems, which should be within the expected uncertainty range. The two comparisons demonstrate the fidelity of our computational approach.

**S3: Slope of the defect formation energy versus the O 2*p*-band center**

We propose that the slope (given in Table S4) of the formation energy of O point defects versus the O 2*p*-band center should correlate to the number of electrons involved in the redox reaction of the defects. For vacancies, which donate two electrons and therefore should ideally have a slope of -2, the actual slopes from our calculations range from -2.05 to -1.60, which is qualitatively consistent with our simple model. For oxide interstitials, which take two electrons, and therefore ideally should have a slope of +2, the actual slopes range from 0.94 to 1.75 depending on defect concentration δ, which is somewhat lower than expected. It seems that the slopes of oxide interstitials decreases with O δ, but those of O vacancies are largely constant. We hypothesize that the increasing 2*p*-band center, which leads to a more covalent material, allows better screening of the defect, which lowers their interactions and thereby reduces their slope when it is positive. This effect would be expected to be stronger for the defects that interact more strongly, which may explain why it occurs for the oxide interstitials but not the vacancies. If this hypothesis is correct then the slope would also be expected to approach +2 for the oxide interstitials at very small δ. Indeed, the slope value for our smallest δ = 0.0625 is 1.75, which is again in reasonable agreement with the implications of our hypothesis.

**Table S1.** Formation energy (eV/ O defect) for O point defects in $(La_{1-x}Sr_x)_2MO_{4\pm\delta}$. The referenced O chemical potential corresponds to just the corrected $O_2$ molecule total energy $\mu_{O_2}^{DFT}$ = -4.38 eV/(O atom)[16].

| M | $\delta$ $x$ | 0.0625 | | | | 0.125 | | | | 0.25 | | | |
|---|---|---|---|---|---|---|---|---|---|---|---|---|---|
| | | Oxi. Int. | Per. Int. | Api. Vac. | Equ. Vac. | Oxi. Int. | Per. Int. | Api. Vac. | Equ. Vac. | Oxi. Int. | Per. Int. | Api. Vac. | Equ. Vac. |
| Co | 0 | -4.0 | -0.6 | 6.5 | 4.5 | -2.7 | -0.1 | 6.7 | 5.4 | -1.6 | 0.0 | 6.4 | 5.3 |
| | 0.25 | -0.6 | -0.1 | 3.8 | 3.3 | -0.4 | 0.0 | 4.1 | 3.3 | -0.3 | 0.1 | 4.8 | 3.9 |
| | 0.5 | 0.2 | 0.1 | 2.8 | 2.4 | 0.1 | 0.0 | 2.5 | 2.5 | 0.5 | 0.2 | 3.1 | 3.1 |
| | 0.75 | 1.4 | 0.1 | 1.7 | 1.3 | 1.3 | 0.3 | 2.2 | 2.0 | 1.4 | 0.4 | 2.6 | 2.2 |
| | 1 | 1.4 | 0.1 | 1.3 | 0.8 | 1.1 | -0.1 | 1.2 | 0.7 | 1.3 | 0.3 | 1.7 | 1.2 |
| Ni | 0 | -2.4 | -0.4 | 6.4 | 5.2 | -1.4 | -0.1 | 6.6 | 5.5 | -0.4 | 0.0 | 6.5 | 5.6 |
| | 0.25 | 0.4 | -0.1 | 2.5 | 2.0 | 0.6 | 0.0 | 2.8 | 2.2 | 0.8 | 0.2 | 3.9 | 3.1 |
| | 0.5 | 1.0 | 0.1 | 1.4 | 1.4 | 1.0 | 0.1 | 1.5 | 1.5 | 1.1 | 0.2 | 1.9 | 1.9 |
| | 0.75 | 1.3 | 0.0 | 0.9 | 0.7 | 1.3 | 0.1 | 1.0 | 0.8 | 1.3 | 0.2 | 1.5 | 1.3 |
| | 1 | 1.2 | 0.0 | 1.0 | 0.6 | 1.5 | 0.0 | 1.1 | 0.7 | 1.5 | 0.2 | 1.5 | 1.0 |
| Cu | 0 | -1.5 | -1.1 | 2.7 | 2.1 | -0.5 | -0.5 | 4.2 | 3.1 | 0.1 | -0.3 | 5.2 | 3.6 |
| | 0.25 | 0.3 | -0.1 | 1.7 | 1.3 | 0.7 | -0.1 | 1.9 | 1.7 | 0.6 | 0.1 | 2.6 | 2.3 |
| | 0.5 | 0.8 | 0.1 | 1.5 | 1.2 | 1.0 | 0.1 | 1.6 | 1.4 | 1.1 | 0.2 | 2.0 | 1.7 |
| | 0.75 | 1.4 | 0.1 | 0.9 | 0.2 | 1.4 | 0.1 | 1.0 | 0.5 | 1.4 | 0.2 | 1.5 | 1.0 |
| | 1 | 1.3 | 0.1 | 0.8 | -0.1 | 1.4 | 0.1 | 1.0 | 0.1 | 1.4 | 0.2 | 1.4 | 0.5 |

**Table S2. Formation energy for O point defect in $(La_{1-x}Sr_x)_2MO4_{\pm\delta}$ compared between theory and experiment.** The referenced O chemical potentials correspond to $P(O_2) = 0.21$ atm for all the four cases and $T$ given in the table for each case. The experimental references[18-20] and this work use $La_{2-x'}Sr_{x'}MO_{4-y}$ and $(La_{1-x}Sr_x)_2MO4_{\pm\delta}$ as chemical formula respectively, so $x = 0.5x'$ and $\delta = |y|$. Theoretical values are those with $\delta$ closest to the experimental $|y|$ ($\delta = 0.0625, 0.125, 0.0625$ and $0.25$ corresponding to $y = 0.01, 0.13, -0.06, 0.4$ from top to bottom, respectively).

| M | $x$ | $T$ (°C) | $E_{def}$ (eV/O defect) Theory[a] (This work) | Experiment[b] |
|---|---|---|---|---|
| Co | 0.5 | 22 | 2.0 | 2.1 (Ref. 18) |
| Co | 0.75 | 22 | 1.7 | 2.1 (Ref. 18) |
| Ni | 0.05 | 704 | -0.7[c] | -1.1 (Ref. 19) |
| Cu | 0.5 | 700 | 0.5 | 0.7 (Ref. 20) |

[a] $E_{def}$ of the most stable O defect (equatorial vacancy, equatorial vacancy, oxide interstitial, and equatorial vacancy from top to bottom)
[b] Converted from $\Delta H_{ox}$ (enthalpy of oxidation) by keeping/reversing sign for O interstitial/vacancy and changing unit from kJ/(mole $O_2$) to (eV/O defect atom).
[c] From linear interpolation of the calculated values with $x = 0$ and $0.25$.

**Table S3. Sr mole fraction $x$ at the transition between O interstitial and O vacancy as the majority defect in $(La_{1-x}Sr_x)_2MO_{4\pm\delta}$ compared between theory and experiment[21-24].** The referenced O chemical potentials correspond to $T$ and $P(O_2)$ given in the table.

| M | $T$ (°C) | $P(O_2)$ (atm) | Critical $x$ | |
|---|---|---|---|---|
| | | | Theory[a] (This work) | Experiment[b] |
| Co | 1000 | 0.21 | 0.35 | 0.4-0.6 (Ref. 21) |
| Ni | 1300 | 0.21 | 0.15 | 0.1-0.2 (Ref. 22) |
| Cu | 800 | 1 | 0.11 | 0.05-0.08 (Ref. 23) |
| Cu | 1000 | 1 | 0.05 | 0-0.1 (Ref. 24) |

[a]The $x$ where O defect concentration $\delta$ changes from positive to negative.
[b]The $x$ where the formation energy curve of the most stable O interstitial crosses with that of the most stable O vacancy. The formation energy curves are linear interpolations of the actually calculated values for $x = 0, 0.25, 0.5, 0.75$ and $1$.

**Table S4. Slope, intercept and coefficient of determination ($R^2$) for linear fitting of formation energy of O point defects versus O 2$p$-band center (relative to the Fermi level) in bulk $(La_{1-x}Sr_x)_2MO_{4\pm\delta}$ (M=Co, Ni, Cu).** The referenced O chemical potential $\mu(O)$ corresponds to $T = 1000$ K and $P(O_2) = 0.21$ atm[16].

|           | $\delta$ | Oxi. Int. | Per. Int. | Api. Vac. | Equ. Vac. |
|-----------|----------|-----------|-----------|-----------|-----------|
|           | 0.0625   | 1.75      | 0.29      | -1.94     | -1.60     |
| slope     | 0.125    | 1.28      | 0.10      | -2.05     | -1.78     |
|           | 0.25     | 0.94      | 0.13      | -1.90     | -1.66     |
|           | 0.0625   | 5.29      | 1.75      | -3.20     | -3.03     |
| intercept | 0.125    | 4.58      | 1.45      | -3.30     | -3.17     |
|           | 0.25     | 4.05      | 1.65      | -2.49     | -2.51     |
|           | 0.0625   | 0.89      | 0.51      | 0.83      | 0.84      |
| $R^2$     | 0.125    | 0.89      | 0.25      | 0.88      | 0.90      |
|           | 0.25     | 0.88      | 0.43      | 0.86      | 0.90      |

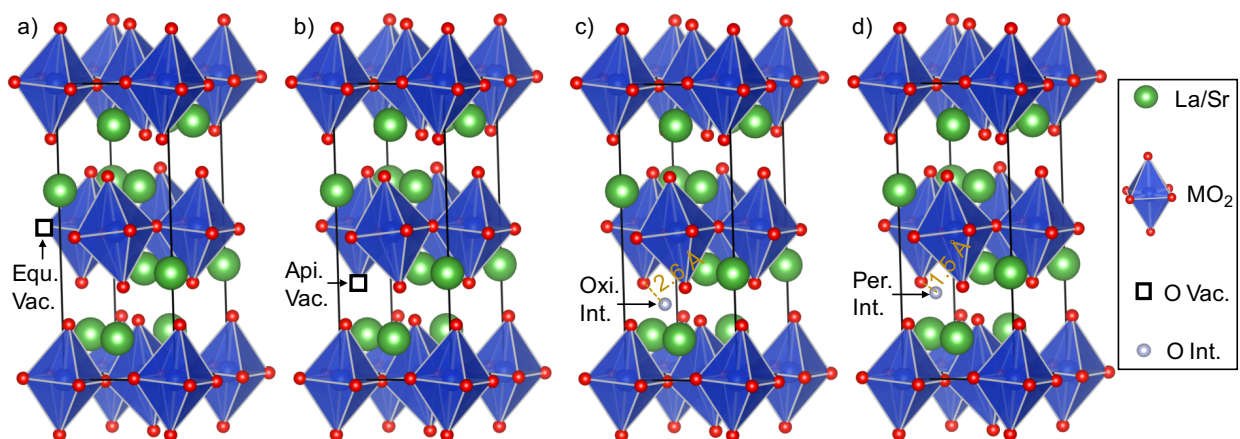

**Figure S1.** Defected supercells of $(La_{1-x}Sr_x)_2MO_{4\pm0.25}$ containing one O point defect of a) equatorial vacancy, b) apical vacancy, c) oxide interstitial, and d) peroxide interstitial.

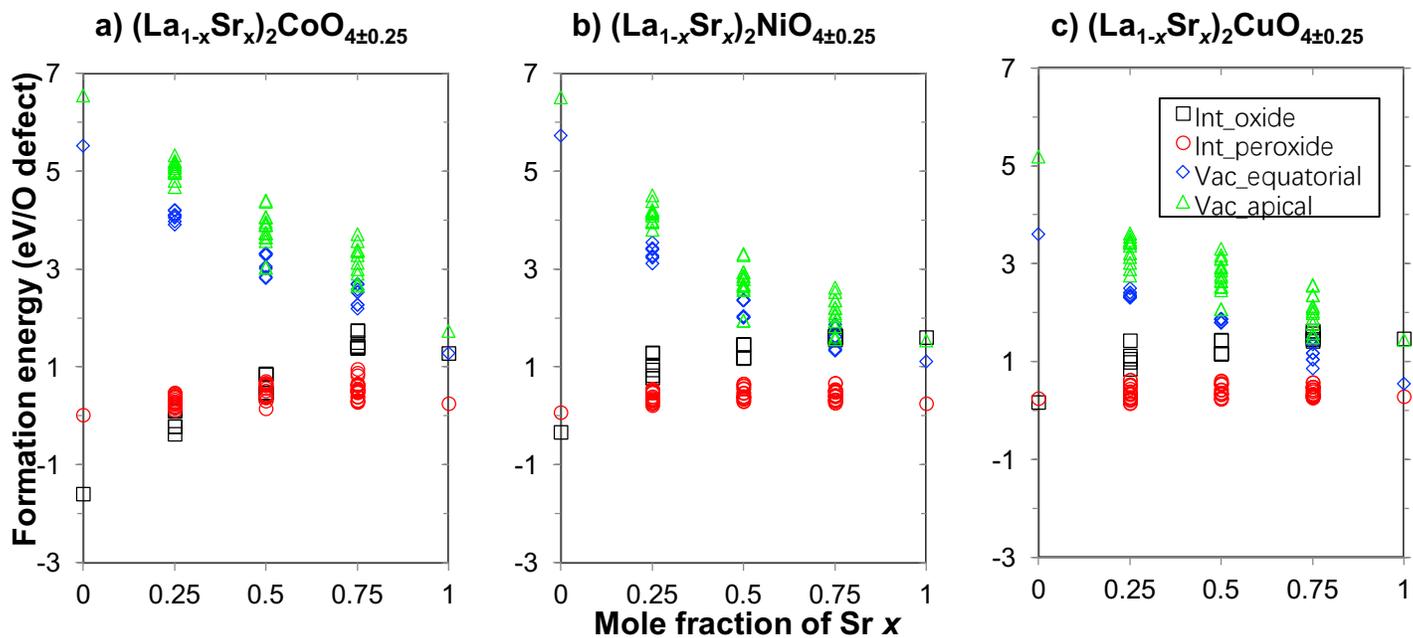

**Figure S2.** Formation energy for O point defects versus Sr mole fraction $x$ in a) $(La_{1-x}Sr_x)_2CoO_{4\pm0.25}$, b) $(La_{1-x}Sr_x)_2NiO_{4\pm0.25}$ and c) $(La_{1-x}Sr_x)_2CuO_{4\pm0.25}$. The referenced O chemical potential corresponds to just the corrected $O_2$ molecule total energy $\mu_{O_2}^{DFT}$ = -4.38 eV/(O atom)[16]. The spread in energy is due to different La/Sr arrangements and symmetry distinct O point defect locations.

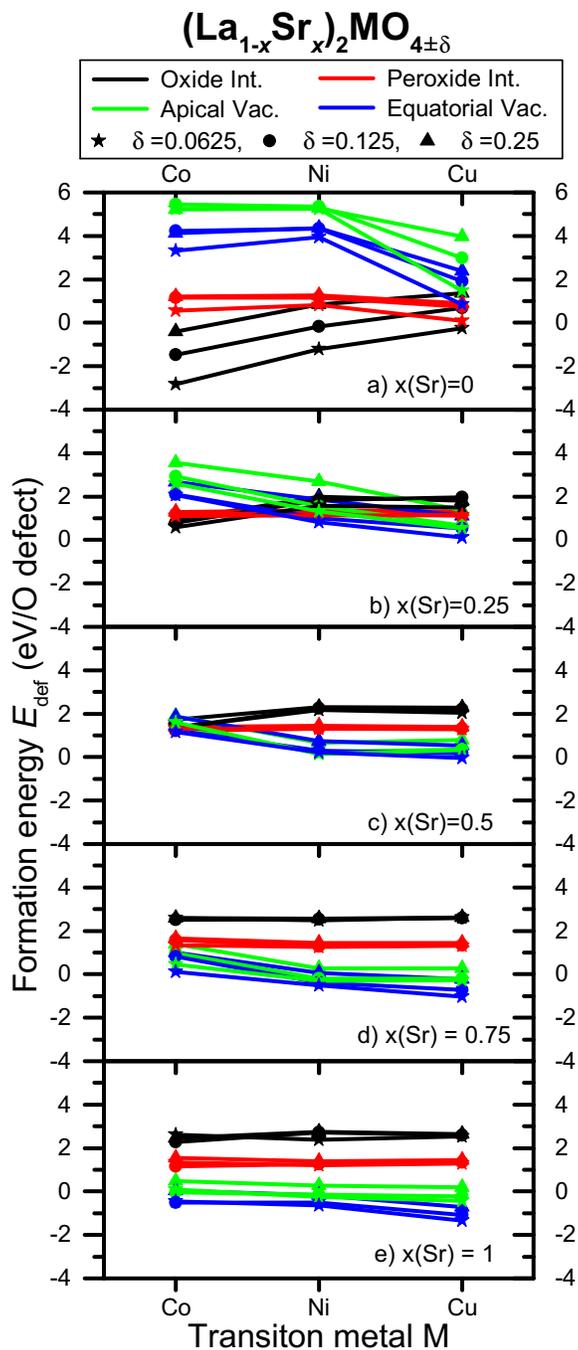

**Figure S3. Formation energy for O point defects versus transition metal M in (La$_{1-x}$Sr$_x$)$_2$MO4$_{\pm\delta}$ with Sr mole fraction $x$ = a) 0, b) 0.25, c) 0.5, d) 0.75, and e) 1.** The referenced O chemical potential corresponds to $T$ = 1000 K and $P(O_2)$ = 0.21 atm[16]. For each type (color) of defect, star, circle and triangle symbols indicate $\delta$ = 0.0625, 0.125 and 0.25, respectively.

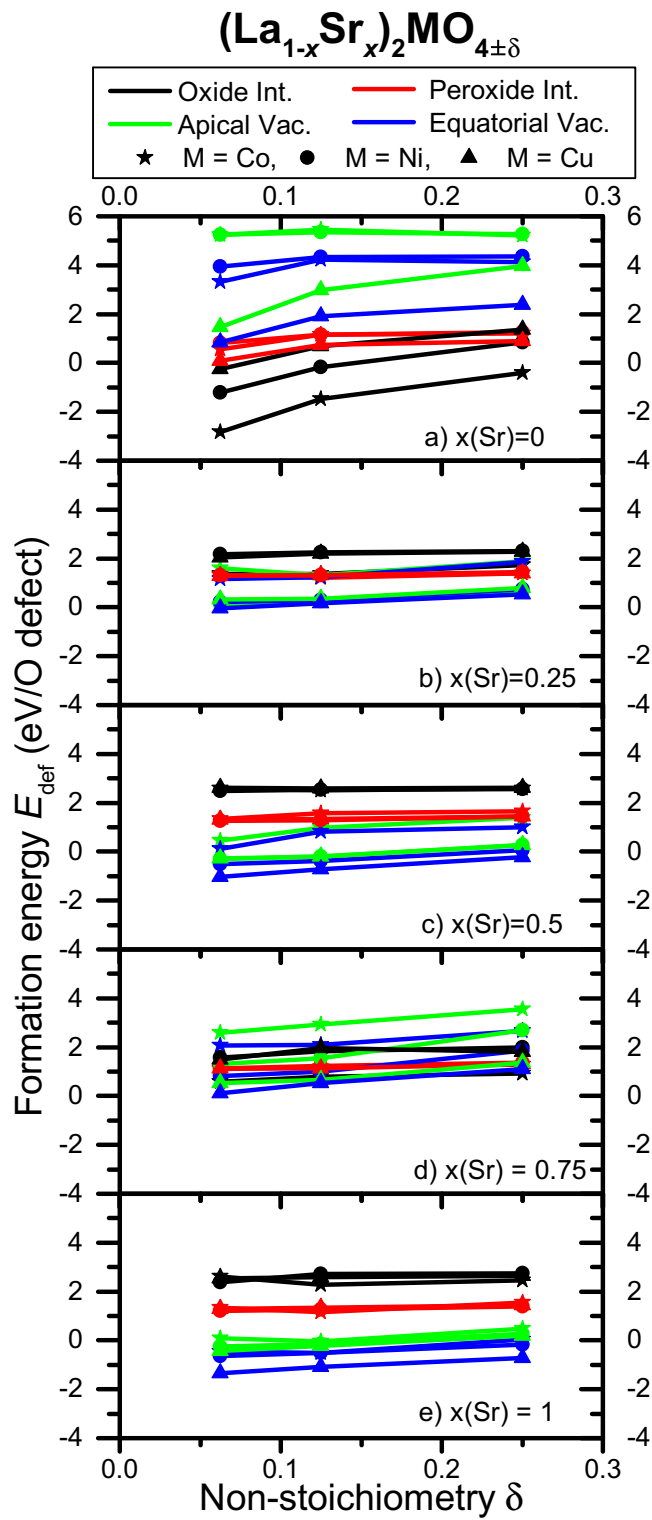

**Figure S4.** Formation energy for O point defects versus O defect concentration δ in (La$_{1-x}$Sr$_x$)$_2$MO4$_{\pm\delta}$ with Sr mole fraction $x$ = a) 0, b) 0.25, c) 0.5, d) 0.75, and e) 1. The referenced O

chemical potential corresponds to $T = 1000$ K and $P(O_2) = 0.21$ atm[16]. For each type (color) of defect, star, circle and triangle symbols indicate M = Co, Ni and Cu, respectively.

# REFERENCES

(1).    Zunger, A.; Wei, S. H.; Ferreira, L. G.; Bernard, J. E. Special Quasirandom Structures. *Phys. Rev. Lett.* **1990,** *65*, 353-356.
(2).    van de Walle, A.; Asta, M.; Ceder, G. The Alloy Theoretic Automated Toolkit: A User Guide. *Calphad* **2002,** *26*, 539-553.
(3).    van de Walle, A. Multicomponent Multisublattice Alloys, Nonconfigurational Entropy and Other Additions to the Alloy Theoretic Automated Toolkit. *Calphad* **2009,** *33*, 266-278.
(4).    Hohenberg, P.; Kohn, W. Inhomogeneous Electron Gas. *Phys. Rev. B* **1964,** *136*, 864-871.
(5).    Kohn, W.; Sham, L. J. Self-Consistent Equations Including Exchange and Correlation Effects. *Phys. Rev.* **1965,** *140*, 1133-1138.
(6).    Kresse, G.; Hafner, J. Ab initio Molecular Dynamics for Liquid Metals. *Phys. Rev. B* **1993,** *47*, 558-561.
(7).    Kresse, G.; Furthmuller, J. Efficient Iterative Schemes for ab initio Total-Energy Calculations Using a Plane-Wave Basis Set. *Phys. Rev. B* **1996,** *54*, 11169-11186.
(8).    Blochl, P. E. Projector Augmented-Wave Method. *Phys. Rev. B* **1994,** *50*, 17953-17979.
(9).    Kresse, G.; Joubert, D. From Ultrasoft Pseudopotentials to the Projector Augmented-Wave Method. *Phys. Rev. B* **1999,** *59*, 1758-1775.
(10).   Perdew, J. P.; Wang, Y. Accurate and Simple Analytic Representation of the Electron-Gas Correlation-Energy. *Phys. Rev. B* **1992,** *45*, 13244-13249.
(11).   Dudarev, S. L.; Botton, G. A.; Savrasov, S. Y.; Humphreys, C. J.; Sutton, A. P. Electron-Energy-Loss Spectra and the Structural Stability of Nickel Oxide: an LSDA+U Study. *Phys. Rev. B* **1998,** *57*, 1505-1509.
(12).   Wang, L.; Maxisch, T.; Ceder, G. Oxidation Energies of Transition Metal Oxides within the GGA+U Framework. *Physical Review B* **2006,** *73*, 195107.
(13).   Davidson, E. R. Matrix Eigenvector Methods, In *Methods in Computational Molecular Physics*, Diercksen, G. H. F.; Wilson, S., Eds. Springer Netherlands: Bad Windsheim, West Germany, 1982; pp 95-113.
(14).   Teter, M. P.; Payne, M. C.; Allan, D. C. Solution of Schrodinger-Equation for Large Systems. *Phys. Rev. B* **1989,** *40*, 12255-12263.
(15).   Monkhorst, H. J.; Pack, J. D. Special Points for Brillouin-Zone Integrations. *Phys. Rev. B* **1976,** *13*, 5188-5192.
(16).   Lee, Y. L.; Kleis, J.; Rossmeisl, J.; Morgan, D. Ab initio Energetics of LaBO3(001) (B=Mn, Fe, Co, and Ni) for Solid Oxide Fuel Cell Cathodes. *Phys. Rev. B* **2009,** *80*, 224101.
(17).   Gurvich, l. V.; Iorish, V. S.; Yungman, V. S.; Dorofeeva, O. V. Thermodynamic Properties as a Function of Temperature. In *CRC handbook of chemistry and physics*, 96th ed.; Haynes, W. M., Ed. CRC Press/Taylor and Francis: Boca Raton, FL, 2015; pp 5-62.
(18).   Prasanna, T. R. S.; Navrotsky, A. Energetics of La2-xSrxCoO4-Y (0.5-Less-Than-x-Less-Than-1.5). *J. Solid State Chem.* **1994,** *112*, 192-195.
(19).   Dicarlo, J.; Mehta, A.; Banschick, D.; Navrotsky, A. The Energetics of La2-xAxNiO4-Y (A = Ba, Sr). *J. Solid State Chem.* **1993,** *103*, 186-192.
(20).   Bularzik, J.; Navrotsky, A.; Dicarlo, J.; Bringley, J.; Scott, B.; Trail, S. Energetics of La2-XSrxCuO4-Y Solid-Solutions (0.0-Less-Than-or-Equal-to-x-Less-Than-or-Equal-to-1.0). *J. Solid State Chem.* **1991,** *93*, 418-429.

(21). Nitadori, T.; Muramatsu, M.; Misono, M. Valence Control, Reactivity of Oxygen, and Catalytic Activity of Lanthanum Strontium Cobalt Oxide (La2-xSrxCoO4). *Chem. Mater.* **1989,** *1*, 215-220.

(22). Takeda, Y.; Kanno, R.; Sakano, M.; Yamamoto, O.; Takano, M.; Bando, Y.; Akinaga, H.; Takita, K.; Goodenough, J. B. Crystal-Chemistry and Physical-Properties of La2-xSrxNiO4(0 <= x <= 1.6). *Mater. Res. Bull.* **1990,** *25*, 293-306.

(23). Kanai, H.; Mizusaki, J.; Tagawa, H.; Hoshiyama, S.; Hirano, K.; Fujita, K.; Tezuka, M.; Hashimoto, T. Defect Chemistry of La2-xSrxCuO4-delta: Oxygen Nonstoichiometry and Thermodynamic Stability. *J. Solid State Chem.* **1997,** *131*, 150-159.

(24). Opila, E. J.; Tuller, H. L. Thermogravimetric Analysis and Defect Models of the Oxygen Nonstoichiometry in La2-xSrxCuO4-Y. *J. Am. Ceram. Soc.* **1994,** *77*, 2727-2737.